# Chemically Tailoring Semiconducting Two-Dimensional Transition Metal Dichalcogenides and Black Phosphorus

Christopher R. Ryder[1†], Joshua D. Wood[1†], Spencer A. Wells[1], and Mark C. Hersam[1,2,3,4*]

[1]*Dept. of Materials Science and Engineering, Northwestern University, Evanston, IL 60208*
[2]*Dept. of Chemistry, Northwestern University, Evanston, IL 60208*
[3]*Dept. of Medicine, Northwestern University, Evanston, IL 60208*
[4]*Dept. of Electrical Engineering and Computer Science, Northwestern University, Evanston, IL 60208*

**ABSTRACT:** Two-dimensional (2D) semiconducting transition metal dichalcogenides (TMDCs) and black phosphorus (BP) have beneficial electronic, optical, and physical properties at the few-layer limit. As atomically thin materials, 2D TMDCs and BP are highly sensitive to their environment and chemical modification, resulting in a strong dependence of their properties on substrate effects, intrinsic defects, and extrinsic adsorbates. Furthermore, the integration of 2D semiconductors into electronic and optoelectronic devices introduces unique challenges at metal-semiconductor and dielectric-semiconductor interfaces. Here, we review emerging efforts to understand and exploit chemical effects to influence the properties of 2D TMDCs and BP. In some cases, surface chemistry leads to significant degradation, thus necessitating the development of robust passivation schemes. On the other hand, appropriately designed chemical modification can be used to beneficially tailor electronic properties, such as controlling doping levels and charge carrier concentrations. Overall, chemical methods allow substantial tunability of the properties of 2D TMDCs and BP, thereby enabling significant future opportunities to optimize performance for device applications.

[*] Correspondence should be addressed to [m-hersam@northwestern.edu](m-hersam@northwestern.edu)

[†] These authors contributed equally.

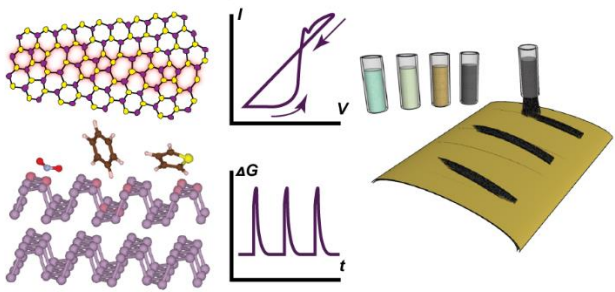

**TOC Figure**

**VOCABULARY: ambipolar –** semiconductor capable of appreciable hole and electron transport, typically controlled by electrostatic gating; **unipolar –** semiconductor that exhibits only electron-dominated (*n*-type) or hole-dominated (*p*-type) transport; **Schottky barrier –** energy barrier for charge carrier transport formed at metal-semiconductor interfaces; **CMOS –** complementary metal-oxide-semiconductor methodology that employs *p*-type and *n*-type transistors for low-power logic operations.

Two-dimensional (2D) nanomaterials have burgeoned into one of the most active fields within nanoscale science and technology over the past decade.[1] Their atomically thin nature provides unprecedented access to pure 2D systems, representing the ultimate thickness scaling limit for semiconductor channels[2] and enabling novel gate-tunable device behavior.[3] Many of the early investigations of chemistry in the 2D limit focused on graphene, including efforts to open an electronic band gap and control interfaces with other materials. These investigations demonstrated how the electronic, optical, and physical properties of graphene can be manipulated by non-covalent and covalent chemical modification,[4-10] as well as how new 2D stoichiometric derivatives can be realized through chemical design.[11]

The more recent emergence of "post-graphene" 2D semiconducting nanomaterials, most prominently transition metal dichalcogenides (TMDCs) and black phosphorus (BP), present other challenges that must be considered if they are to be implemented in high-performance electronic and optoelectronic applications like thin-film transistors, aggressively scaled field-effect transistors (FETs), flexible electronics, complementary metal-oxide-semiconductor (CMOS) logic, light emitters, photodetectors, and light harvesting devices.[12, 13] Here, we review the relationship between the chemistry and semiconducting properties of 2D TMDCs and BP and how desirable attributes may be preserved and engineered by chemical modification.[14] Specifically, we consider the roles of unintentional impurities, ambient adsorbates, and defects, as well as the intentional introduction of surface modification and deviations from stoichiometric, crystalline structure through chemical functionalization schemes. In addition to delineating and categorizing results that have already appeared in the literature, this article critically evaluates the remaining challenges and future prospects for chemically tailored 2D semiconducting nanomaterials.

TMDCs have diverse properties including a wide range of electronic band structures that depend on crystal symmetry and stoichiometric identity.[15] Since semiconducting molybdenum and tungsten TMDCs have band gap energies that span the visible and near-infrared portions of the electromagnetic spectrum, they are particularly promising candidates for electronic and optoelectronic applications.[3, 16, 17] TMDC synthesis has advanced to a state that monolayers can be grown homogeneously at the wafer-scale with metal-organic chemical vapor deposition (MOCVD),[18] while smaller-scale monolayer crystals are routinely prepared on a diverse range of substrates by chemical vapor deposition (CVD) methods.[15] TMDCs FETs can achieve room temperature charge carrier mobilities in excess of ~100 $cm^2$ $V^{-1}$ $s^{-1}$ with large on/off current ratios up to ~$10^8$.[3, 16] $MoS_2$ is the prototypical and most studied TMDC in both electronic and optoelectronic applications. Many chemical functionalization schemes investigated on other nanomaterials have been applied to $MoS_2$, providing a useful reference point for how 2D semiconducting systems can be manipulated chemically. While the other common semiconducting TMDCs based on the transition metals Mo and W and the chalcogens S, Se, and Te have not been investigated as extensively, they are likely to share similar property trends to $MoS_2$.

Unlike the TMDCs, BP is an elemental 2D semiconductor with a buckled, anisotropic crystal structure.[19] With the advent of micromechanical exfoliation methods, interest in the semiconducting properties of BP[20, 21] has renewed recently,[22-25] with a particular focus on BP with nanoscale thicknesses. The electronic band gap of BP varies from 0.3 eV in the bulk to ~2 eV at a monolayer,[21-23] with incrementally larger band gap energies obtained with decreasing thickness in the few-layer limit.[26] At all layer numbers, BP band gaps are predicted to be direct in nature,[26] making them well-suited for optoelectronic applications. In addition to its thickness dependence, the band gap of 2D BP can also be modified by strain[27-31] and electric fields.[31, 32] The anisotropic atomic structure of BP further implies anisotropy in its electronic, optical, and thermal properties,[21-23] leading to efforts to exploit anisotropy for BP device applications. Exfoliated few-layer BP FETs can achieve room temperature hole mobilities of ~1000 cm$^2$ V$^{-1}$ s$^{-1}$ with on/off ratios of ~$10^3$–$10^5$ in back-gated device geometries.[22] Nevertheless, excitement over these favorable device metrics has been tempered by the high chemical reactivity of BP, which leads to degradation in ambient conditions. Chemical passivation schemes are thus of particularly high importance for 2D BP, as outlined in more detail below.

**Semiconductor Nanomaterial Challenges**

Although both TMDCs and BP have promising electronic and physical properties at the few-layer limit, numerous challenges exist for these materials that prevent large-scale integration into nanoelectronic applications.[3, 13] For example, few-layer materials possess exceptionally high surface areas, making them highly sensitive to extrinsic factors. Coulomb scattering (Figure 1a) from charged impurities at dielectric interfaces is particularly deleterious to carrier mobility, leading to efforts to utilize dielectrics with low charge trap densities such as hexagonal boron nitride (h-BN).[33-35] Similarly, ambient molecular species (Figure 1b) compromise charge carrier mobility,[36-38] affect the current-voltage characteristics of devices,[37, 39-41] and cause chemical degradation.[42-45] This high environmental sensitivity of 2D semiconductors often results in reduced charge carrier mobilities for devices that are thinner than ~10 nm.[46] While surface effects can be minimized by employing thicker semiconducting channels, screening effects prevent complete penetration of the gate electric field, thereby degrading current modulation.[47-50] These competing effects can potentially be overcome through the development of chemical passivation schemes that allow high charge carrier mobilities in the atomically thin limit, where current modulation is maximized.

Defects such as grain boundaries and point defects (Figure 1c) are also detrimental to electronic properties, which has motivated the growth of highly crystalline 2D semiconductors. For TMDCs, the heteroatomic bonding additionally produces charged defects that can result in doping effects and mid-gap electronic states.[51-58] For example, defects in MoS$_2$ have been shown to have significant influence on device characteristics.[59] These effects are a consequence of the propensity of MoS$_2$ to possess electronically active defects,[51] such as sulfur vacancies (SVs), which induce strong *n*-type doping. In contrast, BP is a monoatomic semiconductor with homoatomic bonds that typically accommodate uncharged defects, resulting in ambipolar transistor characteristics.[22]

The performance characteristics of TMDC and BP FETs are highly constrained by Schottky barrier (SB) formation and Fermi level pinning contact effects (Figure 1d). Nanomaterials are commonly employed for SB transistors, where the modulation of the SB produces changes in conductance.[60] Interfacial chemistry at the metal-semiconductor interface therefore plays a critical role in device characteristics due to Fermi level pinning and tunnel barrier modification.[61, 62] The chemical concepts of dipole formation, charge transfer, atomic orbital interactions, and electron counting have provided a theoretical basis for contact effects that are poorly described by energy level considerations alone.

Chemistry also plays a critical role in doping, which influences the performance of complementary metal-oxide-semiconductor (CMOS) digital electronics and light harvesting/emission applications that require both *n*-type and *p*-type materials. Additionally, spatially varying doping profiles underlie foundational circuit elements such as *p-n* diodes and bipolar junction transistors. Interfacial chemical modification provides further opportunities to fine tune doping in devices, leading to threshold voltage control that can realize enhancement-mode FETs for low-power electronics.[63] The sections below will consider the unique aspects of these semiconductor nanomaterial challenges for TMDCs and BP.

**Transition Metal Dichalcogenide Overview**

TMDCs consist of three-atom-thick layers in a hexagonal lattice, where transition metals occupy one sublattice position in trigonal prismatic coordination between chalcogens occupying the other sublattice position, as shown in Figure 2a. Optical measurements such as Raman spectroscopy[16, 17, 64, 65] and photoluminescence (PL) are facile methods to measure physical properties of TMDCs and to measure changes in properties due to chemical functionalization. Through inspection of the relative position of the Raman modes of TMDCs, movements of the Fermi level from electrostatic[66] or chemical doping[67, 68] can be deduced. Additionally, information regarding the crystal structure phase[69, 70] and alloying composition[71] has been obtained by monitoring their distinct Raman modes.

Much is inferred about the electronic structure of TMDCs from PL measurements. The bulk band gap of $MoS_2$ is 1.2 eV and indirect, whereas monolayer $MoS_2$ possesses a direct band gap of 1.9 eV (Figure 2b),[72, 73] and is consequently of particular interest for optoelectronic applications.[74-76] The composition of the valence and conduction bands, shown in Figure 2c, is mostly composed of Mo $d$-orbitals with some contribution of S $p$-orbitals.[50] The indirect-direct band gap transition originates from the shifting of the interlayer-sensitive linear combination of Mo $d$-orbitals and anti-bonding S $p_z$-orbitals at the $\Gamma$ point in the Brillouin zone, resulting in the valence band maximum (VBM) and conduction band minimum (CBM) moving to the K point for monolayer $MoS_2$ (Figure 2c).[72, 77] Similar indirect-direct band gap transitions, as well as increases in band gap magnitude,[78] have been observed for monolayer $MoSe_2$ (~1.6 eV),[79-81] $MoTe_2$ (~1.1eV),[82] $WS_2$ (2.0 eV),[83-85] and $WSe_2$ (1.6 eV).[83] The exciton binding energies for 2D TMDCs are significant at ~300–500 meV,[84, 86] and the energy separation of the band edges (quasiparticle band gap) is therefore significantly larger.[86, 87] Two additional spectral features are also commonly observed due to valence band splitting of ~100 meV for Mo and ~400 meV for W.  These effects can be attributed to the large spin-orbit coupling that increases with larger chalcogens,[88, 89] and trion recombination, a feature that depends on the Fermi level position. Trion features are displaced ~30 meV from the exciton photoluminescence (PL) peak due to the trion binding energy and the energy required to add an extra carrier from the Fermi level to the exciton.[90] For TMDCs, the transition metals adopt a 4+ oxidation state and the chalcogens a 2– state.[15] The mixed oxidation states in this heteroatomic bonding produces charged defects that can be electronically active.[52] Defects, in particular chalcogen vacancies, strongly affect the electronic and optical properties of TMDCs. Furthermore, they can change the catalytic (hydrodesulfurization) properties of $MoS_2$.[91] For $MoS_2$, SVs lead to strong $n$-type doping of $MoS_2$ and the resulting unipolar $n$-type conduction in $MoS_2$ FETs. Additionally, SVs produce mid-gap defect states that are highly localized (Figure 2d).[38, 52, 92] Charge transport in $MoS_2$ can occur through both these localized states and electronic bands, with localized transport attributed to Mott variable range hopping,[37, 39, 93] nearest-neighbor hopping,[38] and Arrhenius-type conduction, or a combination thereof.[94] Despite the localized nature of these defect states, increasing the number of SVs by sub-stoichiometric sulfurization in CVD growth and annealing in hydrogen (forming $H_2S$ and a SV) results in higher field-effect mobilities[59] and electrical conductivity (Figure 2e),[91] potentially related to the increased carrier concentration.

Dislocations and grain boundaries in TMDCs are typically composed of irregular non-hexagonal ring pair kinks, which also produce mid-gap electronic states. These kinks vary in stoichiometry and can be either transition metal-rich or chalcogen-rich.[51-57] Consequently, they produce a doping effect, which is manifested as enhancement and quenching in $MoS_2$ PL intensity for tilt boundaries (S-rich) and mirror boundaries (Mo-rich), respectively.[53] Charged defects are also capable of migrating in TMDCs, in part due to the heteroatomic nature of bonding where transition metal homoatomic bonding is not energetically favored.[55] The resulting charge displacement in electric fields produces memristive behavior (Figure 2f), particularly when grain boundaries are in contact with a metal electrode. Since monolayer $MoS_2$ is atomically thin, the memristive response can be tuned by a third gate electrode, presenting new opportunities for advanced computing methods such as neuromorphic computing.[95] For these reasons, stoichiometry and charged defects must be carefully considered, motivating ongoing efforts to engineer vacancies and grain boundaries in TMDCs using high-energy particle[96, 97] and photon beams, as well controlled growth conditions during CVD synthesis.[53, 98]

**Chemistry of Transition Metal Dichalcogenides**

TMDC FETs typically have field-effect mobilities of ~1 to 100 cm$^2$ V$^{-1}$ s$^{-1}$ with high on/off ratios of ~$10^6$ to >$10^8$.[3, 16] Charge transport in TMDC FETs is highly influenced by ambient adsorbates, as evidenced by $MoS_2$ devices measured in ambient typically having far reduced electron mobilities compared to measurements in vacuum.[36-38] Additional improvements in mobility can be achieved by annealing devices in vacuum to further reduce the number of adsorbates.[36] Annealing also shifts the threshold voltage to higher negative potential. As such, ambient adsorbates partially compensate the *n*-type doping of SVs[37, 39-41] and may be chemisorbed at those sites.[36, 99] The detrimental effects of adventitious molecules can be minimized by encapsulating the semiconducting TMDC channel with a high-κ dielectric. For example, devices with a top dielectric and gate electrode, both shielding the channel from adsorbates, have achieved electron mobilities in excess of 100 cm$^2$ V$^{-1}$ s$^{-1}$.

TMDC FETs are commonly found to be SB transistors with strong Fermi level pinning effects.[61, 62, 100] For $MoS_2$, scandium is superior to Au,[101, 102] Pt,[103] Ni,[102] and Ti[94, 104, 105] for low resistance contacts.[103] For all of these metals, independent of work function, carrier injection into the valence band is limited, thus yielding *n*-type behavior that is indicative of significant Fermi level pinning at the $MoS_2$ contacts. Hole injection is further impeded in multilayer $MoS_2$, as its lower valence band[72] relative to monolayer $MoS_2$ gives larger SB heights for holes. In contrast, *p*-type $MoS_2$ transistor channels have been realized by using the deep work function of the $MoO_3$ suboxides. In these suboxides, metallic oxygen vacancy defects, which possess a low density of states, enable effective hole injection into the valence band and result in *p*-type behavior in $MoS_2$ FETs (Figure 3a).[106] However, modest adventitious carbon contamination suppresses Ohmic conduction between the $MoO_3$ and the contact metal, requiring *in situ* contact metal and $MoO_3$ evaporation to avoid increased TMDC contact resistance. Additionally, overall carrier injection into both the valence band and conduction band can still be a bottleneck, and therefore SB transistor behavior is obtained. The lowest contact resistances can be achieved through more complicated designs such as phase-engineering $MoS_2$ to the metallic 1T phase underneath the metal contacts[107] or gate-tunable metal-graphene edge contacts.[49] Other TMDCs also have significant SBs. However, less severe Fermi level pinning occurs with $WSe_2$—potentially due to fewer metal-induced gap states—allowing for SBs to be tuned by the metal work function. Consequently, $WSe_2$ transistors can operate in the unipolar *n*-type,[108] unipolar *p*-type[109, 110] or ambipolar transport regimes.[109] Similarly, a range of charge transport characteristics ranging from unipolar (both *n*-type and *p*-type) to ambipolar has been reported for $MoSe_2$,[111-113] $MoTe_2$,[114, 115] and $WS_2$.[116-118] Higher TMDC drive current by contact engineering is needed not only for TMDC-based logic but for also improving device sensitivity to sensor targets.

Controlled doping of TMDCs can be achieved through non-covalent functionalization with redox active molecules and inorganic compounds. In particular, heavy doping lowers the resistance of metal contacts as has been demonstrated with *n*-type doping of $MoS_2$ by polyethlyeneimine,[119] benzyl viologen,[68] Cl (although this may involve Cl substitution at SVs),[120] and K.[121] Similar effects have also been obtained by *n*-type doping $WS_2$ with Cl.[120] That said, halogen and alkali non-covalent TMDC modifications present challenges for conventional semiconductor processing. The contact resistance is also lowered by heavily *p*-type doping $WSe_2$ FETs with Pd contacts[109] and degenerately *n*-type doping $WSe_2$ FETs with Au contacts.[121] A general feature of the aforementioned doping techniques is that the device threshold voltages are shifted to greater negative potential for *n*-type doping and to greater positive potential for *p*-type doping. Furthermore, the reduced contact resistances yield increases in on-current and mobility due to more efficient charge carrier injection and collection.

The resulting doping level can be deduced without device fabrication by monitoring shifts in the PL spectra, as the free carrier density changes the relative amounts of trion and exciton recombination. Tetrafluorotetracyanoquinodimethane ($F_4$-TCNQ) is a strong electron acceptor and effectively *p*-type dopes $MoS_2$ and $WS_2$, increasing exciton recombination and thereby PL emission intensity and photon energy (Figure 3b). Conversely, nicotinamide adenine dinucleotide (NADH) is an electron donor molecule that *n*-type dopes $MoS_2$, resulting in reduced emission from excitons as the concentration of trions increases.[122] The evaporation of $Cs_2CO_3$ films, which also serve as electron donors, *n*-type dope $MoS_2$ and produce similar reductions in PL emission intensity and energy.[123] By masking a portion of the $MoS_2$ FET channel and doping the exposed area with $AuCl_3$, *p-n* junctions have also been fabricated that show enhanced rectification behavior due to the asymmetric contacts.[124] In comparison to non-covalent chemistries, covalent modification of TMDCs are less prevalent. At high annealing temperatures, oxygen produces nanoscale trenches in $MoS_2$ that have strong PL emission intensity (Figure 3c). Similar results have been obtained by oxygen plasma treatments that produce reactive oxygen species.[125] Conversely, enhancements in PL emission efficiency have also been obtained *via* the treatment of TMDCs with the bis(trifluoromethane) sulfonimide[126] and hydrobromic acid,[127] which have been attributed to the repairing of structural or chemical defects. $MoS_2$ and $MoSe_2$ can also react covalently with organic molecules,[128] particularly in conjunction with lithiation treatments. For TMDCs, lithiation creates defects, increases electron concentrations, and drives crystal structure phase changes from the semiconducting 2H phase to the metallic 1T phase,[129] thereby enabling reactions with electrophiles such as organic halides and aryl diazonium compounds (Figure 3d).[130,131] In some cases, this subsequent covalent chemistry results in a reversion from the 1T phase back to the 2H phase, restoring the TMDC semiconducting properties and modifying PL.[130] In addition, $WSe_2$ has been found to incorporate $NO_x$ species at Se positions after annealing at ~150 °C, leading to robust, air-stable *p*-type doping in FETs that persists even following annealing at ~300 °C in inert environments.[132] Indeed, TMDC lithiation is limited by the need for water-free processing, either in a glove box or Schlenk line.

Alloying is another rich area for the modification of TMDC electronic properties, whereby both the transition metals and chalcogens can be substituted, producing ternary alloys with modified VBM and CBM positions. Specifically, molybdenum and tungsten alloys have been grown and exfoliated from bulk crystals. In these cases, the valence band decreases in energy linearly with increasing tungsten composition due to $d_{xy}$ and $d_{x2-y2}$ orbital contributions to the VBM. On the other hand, the conduction band decreases nonlinearly (*i.e.*, "bows") due to the differing orbital symmetries of the CBM in Mo and W.[133] Alloying with Nb is especially advantageous, because Nb has one fewer electron than Mo, resulting in *p*-type doping in $Mo_{1-}$ $_x$Nb$_x$S$_2$ alloys that is substantial enough to produce *p*-type FETs.[134] MoS$_{2(1-x)}$Se$_{2x}$ monolayer alloys have also been grown during CVD by either mixing the S and Se precursors in various concentrations[71] or by using separately heated chalcogen precursor boats.[135] Both methods allow for the composition to be tuned between pure MoS$_2$ and pure MoSe$_2$ with resulting PL spectral control between the respective optical band gaps. The latter method is particularly well-suited for producing homogeneous PL emission across flakes (Figure 3e). Ternary alloys have also been produced through Ar sputtering of MoS$_2$ and MoSe$_2$ followed by diselenodiphenyl or benzenethiol treatment, respectively, albeit with limited ambient stability.[136]

**Black Phosphorus Overview**

Black phosphorus possesses an orthorhombic, buckled crystal structure with each phosphorus atom having two bonds constituting the zigzag direction, one bond in the armchair direction, and a lone pair of electrons (Figure 4a). The two types of bonds along the orthogonal in-plane directions of BP differ in bond length and bond angle, producing anisotropy in the mechanical,[137-139] thermal,[140, 141] electrical,[142, 143] and optical properties.[143-148] The orthorhombic structure of BP makes it unique among the group 15 elements,[149] and arises from *s-p* orbital mixing that is stronger for lighter group elements.[19] In contrast, As, Sb, and Bi all have layered rhombohedral structures in part due to less significant *s-p* orbital mixing, but also due to increased interlayer interactions.[19] BP has also been found to adopt a rhombohedral crystal structure at high pressures ~5 GPa,[150, 151] which was recently theorized to result from the interlayer interactions of electron lone pairs rearranging the bonding structure.[152]

The electron lone pairs of BP require special consideration due to their roles in strong interlayer interactions and chemical reactivity. Bulk BP has a disperse band structure in the out-of-plane direction and consequently low charge carrier effective masses.[21, 153] Moreover, a substantial out-of-plane mobility of $8.5 \times 10^3$ cm$^2$ V$^{-1}$ s$^{-1}$ has been measured at low temperature.[154] Significant charge redistribution between layers is facilitated by the electron lone pairs,[155] and is manifested by interlayer interactions that are more substantial than described by simple van der Waals forces. Evidence of relatively strong interlayer interactions are found in the stronger dependence of the Raman interlayer breathing mode on layer number for BP than TMDCs, as well as a larger interlayer force constant.[147] For these reasons, the electron lone pairs can be considered to be labile and a source of chemical reactivity, as they can redistribute and engage in covalent interactions.

The band gap of BP decreases from ~2 eV for a monolayer to 0.3 eV in the bulk. In contrast to TMDCs, the band gap of BP is direct for all thicknesses (Figure 4b).[26] The band structure of BP shows strong anisotropy along the Γ–X and Γ–Y directions for all thicknesses, with the Γ–X (armchair) direction far more disperse than the Γ–Y (zigzag) direction.[48, 153, 156] This anisotropy leads to significantly different charge carrier effective masses as a functional of crystallographic orientation and *quasi*-1D electronic properties.[155] While there is moderate *s*-orbital character (and perhaps *d*-orbital contributions),[157] the valence and conduction bands are predominantly *p*-orbital derived, with the VBM and CBM comprised of $p_z$-orbitals (Figure 4c).[153, 158-160] The linear combination of these $p_z$ orbitals with additional layers is responsible for the variation in the magnitude of the band gap.[153] The direct nature of the band gaps of bulk and few-layer BP make it highly suited for optoelectronic applications over a broad range of the electromagnetic spectrum. Moreover, by utilizing the anisotropy of BP (Figure 4d), a polarization-sensitive response can be obtained (Figure 4e).[159]

Charge transport in BP FETs is ambipolar,[50, 154, 161, 162] such that the Fermi level can be shifted into both the valence and conduction bands. The ability to electrostatically dope both *n*-type and *p*-type is promising for complementary doping schemes, although enhanced hole transport is often obtained.[21, 22, 154] The band-like transport of BP and its ambipolar characteristics are different than TMDCs, in particular $MoS_2$, which may be due to electronically inactive defect states.[163] BP FETs typically have room temperature hole mobilities that range between 100 and 1,000 cm$^2$ V$^{-1}$ s$^{-1}$,[48, 142, 164] with measurements almost exclusively performed in vacuum. Low temperature transport measurements show that the temperature dependence of the field-effect mobility of BP is weaker than TMDCs, indicating that transport is more robust against phonon scattering.[48] By reducing Coulomb scattering with hexagonal boron nitride as a substrate, enhanced FET performance is obtained at low temperatures (~2 K), with hole and electron mobilities reaching >1000 cm$^2$ V$^{-1}$ s$^{-1}$ and 4000 cm$^2$ V$^{-1}$ s$^{-1}$, respectively.[50, 165, 166] The high carrier mobilities of BP enables high performance in RF applications[167] and photodetectors,[168] with initial BP devices showing comparable, and in some cases, superior performance to analogous early-stage $MoS_2$ and graphene devices.

**Chemistry of Black Phosphorus**

BP is the most stable allotrope of phosphorus at ambient conditions based upon enthalpy measurements,[157] with few-layer BP prepared by mechanical exfoliation thermally stable in vacuum conditions up to ~400 °C.[169] Nevertheless, ambient environmental exposure is notably detrimental to exfoliated few-layer BP. The flat surface of freshly-exfoliated flakes roughens and forms morphological protrusions that are revealed by atomic force microscopy (Figure 5a) and correlated to changes in the electronic and chemical properties of BP.[42-44, 170] For BP FETs, the adsorption of ambient species first results in a shift of the threshold voltage to higher positive voltages, indicative of *p*-type doping,[42-44] as predicted for the physisorption of $O_2$ on BP.[171] This effect can be reversed somewhat following BP device annealing, whereby ambipolar conduction is restored.[172] Nevertheless, extended ambient exposure is more pernicious and ultimately leads to a loss of conduction within two days (Figure 5b).[43] It is plausible that ambient adsorbates (*e.g.*, oxygen and water) first interact non-covalently, resulting in a reversible effect, and then react covalently, irreversibly disrupting molecular orbitals and ultimately compromising electronic properties.

The formation of phosphorus oxide results in structural and chemical changes that can be measured spectroscopically. Raman spectroscopy of BP exposed to ambient conditions reveals a reduction in the intensity of Raman modes.[45, 173, 174] Measurements performed in ambient conditions show a decrease in the integrated $A^1_g$ to $A^2_g$ ratios and an overall reduction in BP Raman modes following hours of laser irradiation (Figure 5c), suggesting photo-oxidation as a mechanism of degradation.[45] Infrared spectroscopy further shows a broad mode at ~880 cm$^{-1}$ due to P–O ester stretching.[43] Electron-energy loss spectroscopy (EELS) and X-ray photoelectron spectroscopy (XPS) measurements of exfoliated flakes have features attributed to phosphorus oxide species between ~134–136 eV.[43, 45] The identity of these species has been assigned by measuring the oxides formed on bulk BP crystals with synchrotron X-ray methods. Small amounts of $P_2O_4$ moieties are responsible for peaks at 131.54 eV (O–P=O) and 132.67 eV (P–O–P), with the spectrum dominated by $P_2O_5$ at 134.65 eV, as shown in Figure 5d.[175]

Density functional theory (DFT) calculations have found that the chemisorption of oxygen is favorable thermodynamically, producing P–O moieties with large binding energies.[176, 177] In addition, the formation of both stoichiometric and nonstoichiometric oxides are found to be stable.[178] The energy barrier for the dissociation of oxygen molecules on BP is relatively large (>

3 eV), and therefore dissociation has been proposed to be assisted by photons[176, 179] and defects.[179] The overall favorable energetics of BP ambient oxidation must be circumvented by BP passivation schemes for technologically relevant applications. Towards that end, BP devices have been encapsulated with dielectric materials grown from ALD,[43, 44, 180, 181] h-BN,[166, 174, 182] which forms atomically clean interfaces,[50] and polymers,[44] thus allowing semiconducting properties to be preserved for weeks or longer in ambient conditions. While the oxide formed on bulk BP has been found to be stable,[175] the diffusion of ambient species from edges and substrate interfaces still can play a significant role in exfoliated flake degradation, making passivation schemes for few-layer BP necessary. The kinetics for edge *versus* surface-introduced BP oxidants are not definitively established, though there are indications that edge-based oxidation is faster than surface-based oxidation. In turn, more reactive BP edges could be passivated chemically, resulting in an active BP surface while preventing wholesale oxidation.

PL from BP has been attributed to the enlarging of the band gap at the few-layer limit. PL emission at ~1.6-1.75 eV (~780–710 nm), ~1.3 eV (~950 nm), ~1.0 eV (~1240 nm), ~0.9 eV (~1380 nm), and ~0.8 eV (~1550 nm) have been assigned to one to five layers of BP (Figure 6a).[50, 142, 173, 183-185] These emissions have been demonstrated to be highly anisotropic (Figure 6b), suggesting that BP may have *quasi*-1D excitons.[185] Nevertheless, the assignment of these peaks has varied and is further complicated by PL peaks being attributed to trion recombination[184, 186] and oxygen defects.[187] Depending on PL assignment, the exciton binding energy in BP can vary by hundreds of meV, assuming that the emission results from exciton-based recombination. However, for BP samples encapsulated in h-BN in inert environments, PL emission has only been measured from BP flakes and regions that have been exposed to ambient conditions, as shown in Figure 6c.[50] This result coupled with the recent report of PL in the visible range of intentionally oxidized bulk BP (Figure 6d),[188] indicate that oxygen may play a role in the optical properties of BP, necessitating further investigation of the chemical origins of light emission.

While TMDCs show FET performance dominated by Schottky barriers at the contacts, it is possible to obtain higher BP channel resistance than contact resistance with Ti[189] and Pd[190] metal contacts. Nevertheless, *p*-type conduction and Schottky barriers most often dominate transistor behavior. In addition to adventitious *p*-type doping in BP FETs, Fermi level pinning likely occurs at the contacts, as hole injection is favored for both small work function Ti and large work function Pd.[161] However, ambipolar and *n*-type channel conduction has been obtained with Al contacts[190] and by exploiting the layer-dependent VBM and CBM positions.[26] Additionally, Cu, Ta, and Nb have been proposed to make effective contacts to monolayer BP.[191] In particular, covalent interactions between Cu and BP have been predicted to be Ohmic in nature and facilitate electron injection. Similarly, electron injection has been found to be possible for Ta and Nb. Like the TMDCs, it may be possible to phase transform BP[151] contact regions, lowering contact resistance *via* BP band structure modifications and covalent chemistry.

The evaporation of $Cs_2CO_3$ and $MoO_3$ effectively modulates the Fermi level of BP and reduces the Schottky barriers at the metal-BP interface, ultimately lowering contact resistances in BP FETs. Additionally, $Cs_2CO_3$ deposition incrementally shifts the threshold voltage to greater negative potentials and enhances electron concentration and mobility, effectively creating an *n*-type device (Figure 7a). $MoO_3$ channel functionalization has similarly been found to shift the threshold voltage to greater positive potentials and increase the hole concentration (Figure 7b).[189] Tellurium impurities can also produce *n*-type conduction in bulk BP,[154] although similar results have not yet been reported in few-layer limit. Finally, potassium *n*-type dopes BP in addition to shifting the VBM and CBM towards each other, forming a topologically insulating state when the bands touch.[192]

Recent theoretical work has examined the adsorption of dilute quantities of small molecules on BP, concluding that CO, $H_2$, $H_2O$, and $NH_3$ are weak electron donors, while $NO_2$, NO, and $O_2$, are electron acceptors.[171] $NO_2$ is predicted to have the largest charge transfer from BP, which is similar to the *p*-type doping induced by $NO_2$ for other 2D nanomaterials.[193] Nevertheless, the volatility of $NO_2$ raises questions about its practicality as a dopant in device applications. Conversely, electron donor and acceptor organic molecules have much lower vapor pressures, allowing them to be cast onto devices with stability in ambient conditions.[68] Computational studies of the adsorption of TTF and TCNQ, common donor and acceptor molecules, respectively, have shown that effective doping with redox active molecules is difficult due to large energy barriers that frustrate large charge transfer.[171]

Despite the inherent reactivity of BP, relatively few covalent functionalization schemes have been demonstrated. Phosphorus is capable of adopting higher coordination numbers than what is present in BP. Phosphorus is well-known to form bonds with N, C, O, and the halogens, as well as Lewis pairs through lone pair interactions with Lewis acids.[194] Therefore, it is possible to expect Lewis pairing between BP lone pairs and small alkali ion, transition metal ion, and enone-based Lewis acids. Stable phosphorus-carbon bonds between bulk BP and graphite have been created through high-energy ball milling, creating composite materials for Li-ion battery anodes. These bonds were created at the edge interfaces of BP and graphite.[195] Theoretical predictions of gas adsorption have found that nitric oxide chemisorbs on BP *via* the dangling bond of unpaired electrons.[196] Additionally, chemical functionalization with hydrogen, fluorine, and imine groups results in P–H, P–F,[197] and P=N bonds[177] that modify the electronic structure of BP (Figure 7d).

Alloying with arsenic presents a final chemical method for tuning the properties of BP. Black arsenic-phosphorus alloys have been prepared by sealing a combination of gray (trigonal) arsenic and red phosphorus with a mineralization agent in a glass ampoule and heating to temperatures in excess of 500 °C.[198] Depending on the mineralization agent used, arsenic compositions as high as 83% have been obtained in the orthorhombic structure, resulting in distinctive broad arsenic-arsenic Raman modes at wavenumbers below 300 cm$^{-1}$ and phosphorus-arsenic Raman modes between 300 and 380 cm$^{-1}$. Arsenic alloying allows for the band gap to be tuned downward to ~0.15 eV (~8270 nm), as measured by infrared absorption (Figure 7c).[199] Due to the relatively few viable semiconductors with such narrow gaps, black arsenic-phosphorus alloys are promising alternatives to existing HgCdTe technologies.[200]

**Future Outlook**

The emergence of a new semiconductor inevitably leads to efforts to consider it as a potential replacement for silicon in conventional integrated circuit technology. Indeed, recent efforts to improve TMDC growth by CVD are using FET mobility and current on/off ratio as process benchmarks[18, 53, 54, 59] with the principal aim of synthesizing monocrystalline (>100 μm) and chemically homogeneous TMDCs for large-scale integration. Although improvements in material quality will undoubtedly have value, history is littered with failed attempts to supplant silicon in conventional electronics, and therefore applications that are uniquely enabled by 2D semiconductors should be explored in parallel. For example, a recent report[95] has demonstrated that small-grain, sub-stoichiometric MoS$_2$ can reach unprecedented regimes of non-linear charge transport, including gate-tunable memristive phenomena. In contrast to existing two-terminal memristor technologies, the 2D TMDC memristive response can be controlled by a third gate terminal, which is promising for emerging efforts in neuromorphic computing. Chemical modification is likely to play a significant role in these efforts, as 2D TMDC memristive characteristics are strongly influenced by grain boundaries and stoichiometry. Consequently, engineering both the chalcogen vacancy density and grain boundary physical and chemical structure should enable tuning of the memristive response (Figure 8a). The incomplete screening of vertically applied electric fields that results from the atomically thin nature of 2D semiconductors also allows other conventional two-terminal devices to be modulated by a third gate terminal, creating additional opportunities for non-FET devices. For example, $MoS_2$ has recently been employed as the *n*-type semiconductor in gate-tunable *p-n* diodes[201, 202] that show unusual anti-ambipolar responses. Such devices are well-suited as foundational elements of communications technologies including frequency doublers and phase/frequency shift keying circuits.[203]

Since exfoliated TMDC samples have been available for a longer period of time than exfoliated BP, it is likely that many of the early chemical modification attempts for BP will mirror efforts already pursued for TMDCs. Nevertheless, the higher chemical reactivity of BP[43] should facilitate additional opportunities for chemistry beyond those that have been demonstrated for TMDCs (Figure 8b). For example, the lone pairs on the surface of BP may enable the development of novel 2D covalent—and possibly non-covalent—interactions. While ambient BP oxidation makes controlled BP chemistry challenging, plentiful opportunities exist to process BP in inert environments (*e.g.*, glove box, Schlenk line, and vacuum) or by avoiding the oxidizing combination of $H_2O$, $O_2$, and light.[43, 45] Vacuum-phase, atomic O, H, N, and F functionalization[9] of BP may drive surface and interlayer chemistry, forming sub-stoichiometric and stoichiometric BP derivatives akin to graphene oxide,[5] graphane,[204] and fluorinated graphene.[205] An initial report of BP non-covalent functionalization by $Cs_2CO_3$ and $MoO_3$ has demonstrated encouraging manipulation of BP FET transport and metrics,[189] thereby motivating further exploration of non-covalent modification chemistries for BP. Non-covalent BP modification should enhance the sensitivity and selectivity of BP-based sensors (Figure 8b) that have already shown sensitivities down to the part per billion level for $NO_2$.[206, 207]

The established arsenic alloying of BP suggests that other group 15 elements such as N, Sb, and Bi can also be alloyed in the orthorhombic BP lattice (Figure 8c). In addition, chalcogens (*e.g.*, Te, Se, and S) may prove to be effective interstitial or substitutional dopants for 2D BP, as was demonstrated for Te in BP bulk crystals.[154] During As alloying of BP, it is also possible that orthorhombic BP "seeds" may nucleate layered structures that are P-free, facilitating the production of black arsenic, an orthorhombic cousin of BP.[198, 208] Such a seeded approach may allow the synthesis of BP isostructures of 2D As, Sb, and Bi[209] (Figure 8d) that have not yet been realized experimentally,[210] with the exception of 2D Bi allotropes enabled by strong Si(111)–7×7 substrate interactions.[211]

Heteroelemental 2D isostructures of BP also exist (Figure 8d), employing elements from groups 14 and 16 (IV–VI). Among these potential isostructures, 2D SnS,[212, 213] SnSe,[214] GeS,[215] and GeSe[215, 216] have been isolated, showing structural perturbations from a starting rocksalt configuration that are analogous to BP.[217, 218] Other IV–VI compound semiconductors such as SiS[219] have been theorized, morphologically deviating from orthorhombic BP. In the bulk, the group 14 monochalcogenides SnS, SnSe, and GeSe all possess near-infrared (<1.45 eV) indirect band gaps, whereas GeS has a 1.65 eV direct band gap.[220] With anisotropic atomic structure, 2D SnS, SnSe, GeS, and GeSe are expected to exhibit similar in-plane anisotropy as BP (Figure 8e[221]), leading to unusual exciton physics[222] and opportunities in energy conversion technologies like thermoelectrics.[223] Single crystal SnSe has been demonstrated to have record thermoelectric performance, achieved by increasing the conductivity and Seebeck coefficient through $p$-type doping with Na.[224] The heteroelemental, heavy atoms in 2D SnS, SnSe, GeS, and GeSe, coupled with the buckled BP structure, should result in strong spin-orbit coupling, leading to valley-dependent[225] and topological effects.[226] Finally, strategies for modifying TMDCs chemically by chalcogen addition, defect engineering, or surface restructuring may be needed for the group 14 monochalcogenides, since the as-grown monochalcogenide compounds often have non-stoichiometric impurities that impact their properties.

As the most promising applications for 2D TMDCs and BP become apparent, the need for large-scale production of homogeneous 2D TMDC and BP samples will increase. For the TMDCs, scale-up efforts have already commenced, including solution-based exfoliation and thickness sorting by density gradient ultracentrifugation,[227] in addition to the aforementioned wafer-scale MOCVD growth methodology.[18] On the other hand, BP growth by CVD or molecular beam epitaxy (MBE) likely requires precursors like phosphine and organophosphorus derivatives, whose high toxicity will necessitate costly investments in safe gas handling equipment and careful understanding of chemical side reactions. To avoid these complications, recent work has focused on the growth of BP thin films by pulsed laser deposition[228] and phase transformation of deposited red phosphorus,[229] but the electrical performance of these BP thin films has thus far fallen well short of micromechanically exfoliated BP. In contrast, BP nanosheets have been exfoliated in anhydrous organic solvents (Figure 8f),[230-233] with the resulting electrical and chemical properties competing favorably with micromechanically exfoliated BP. Further improvements in sample quantity and quality therefore appear inevitable, accelerating the ongoing fundamental studies and anticipated applied technologies for chemically tailored 2D TMDCs and BP.

**Conflict of Interest:** The authors declare no competing financial interests.

**Acknowledgments**

This work was supported by the Office of Naval Research (ONR N00014-14-1-0669) and the Materials Research Science and Engineering Center (MRSEC) of Northwestern University (National Science Foundation Grant DMR-1121262). S.A.W. was supported under contract FA9550-11-C-0028 from the Department of Defense, Air Force Office of Scientific Research, National Defense Science and Engineering Graduate (NDSEG) Fellowship.

**Figures**

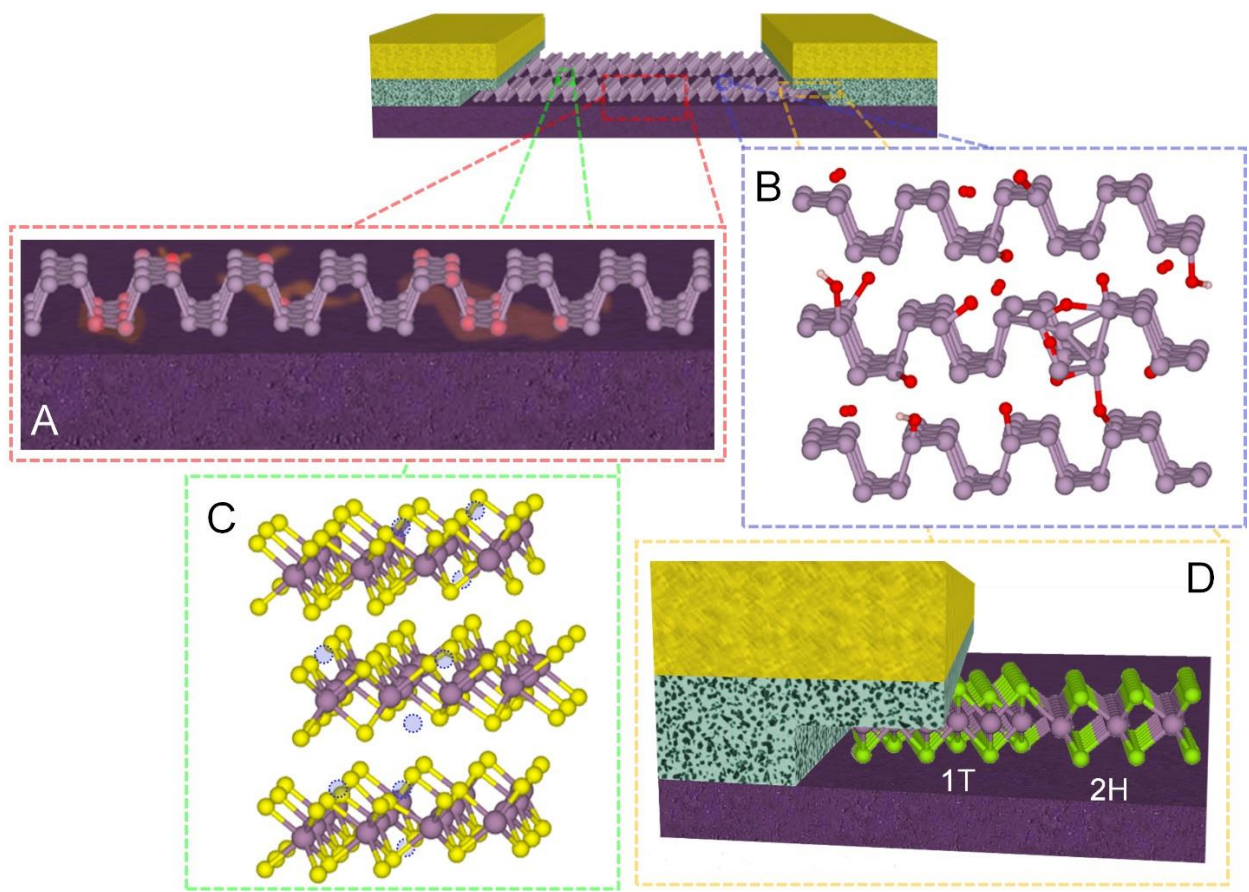

**Figure 1. Chemical considerations for semiconducting 2D nanomaterial devices. (A)** Substrate and top dielectrics for nanomaterials can serve as sources and barriers for charge impurities that produce carrier scattering. **(B)** Extrinsic adsorbates, unintentional and intentional, can dominate the electronic and optical properties of 2D nanomaterials by shifting the Fermi level and, in the case of chemisorption, by changing the electronic band structure. **(C)** Defects can be sources of carrier scattering and can dominate the charge transport mechanism by producing localized electronic states. **(D)** Contact effects such as Schottky barriers can limit carrier densities and influence carrier mobility.

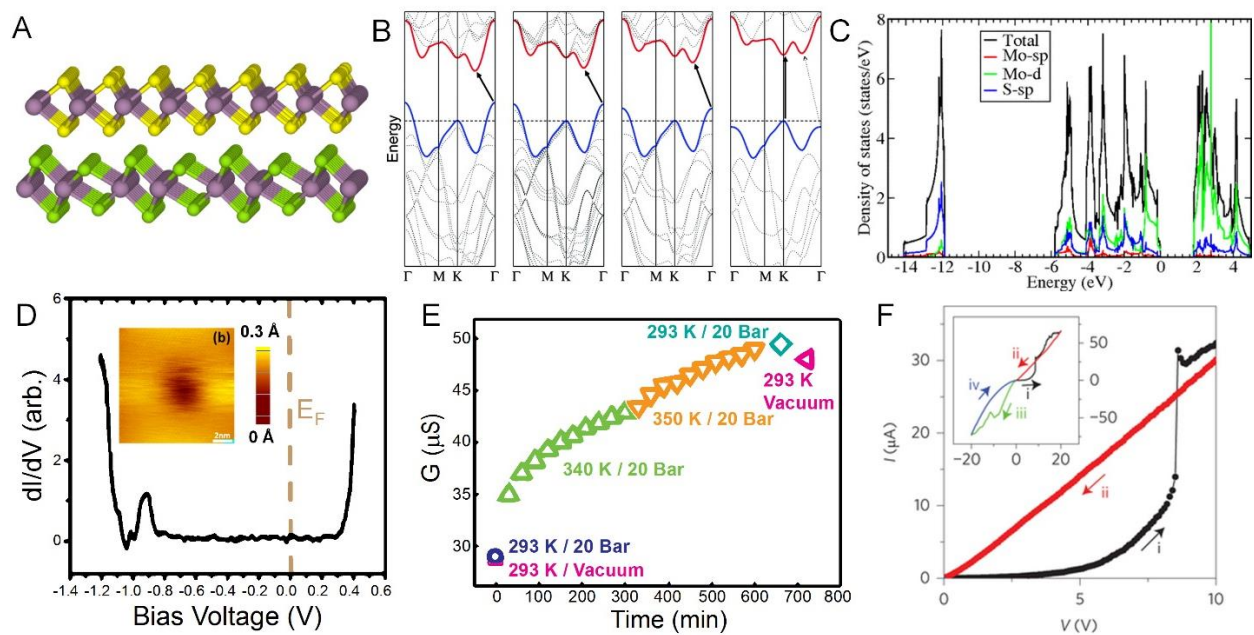

**Figure 2. TMDC crystal structure, electronic structure, and defects. (A)** Crystal structures of 2H-MoS$_2$ and 1T-MoSe$_2$. **(B)** Band structures of bulk, trilayer, bilayer, and monolayer MoS$_2$ (from left to right) calculated by DFT. These calculations show the increase in band gap and the transition from indirect band gap to direct band gap for monolayer MoS$_2$. Adapted with permission from ref 72. Copyright 2010 American Chemical Society. **(C)** Projected density of states of MoS$_2$. The VBM is composed predominantly of Mo *d*-orbital character, while the CBM is composed of Mo *d*-orbital and S *s*-orbitals. Adapted with permission from ref 234. Copyright 2009 American Physical Society. **(D)** Scanning tunneling spectroscopy *dI/dV* spectrum of a sulfur vacancy in MoS$_2$ (constant height scanning tunneling microscopy image in the inset) showing a distinct midgap state. Adapted with permission from ref 92. Copyright 2014 American Chemical Society. **(E)** Enhancement in conductance of MoS$_2$ with the creation of sulfur vacancies following annealing at 67 ˚C in 20 Bar of H$_2$. The resulting increased conductance persists at room temperature in the absence of H$_2$ gas. Adapted with permission from ref 91. Copyright 2013 The Royal Society of Chemistry. **(F)** Memristive device characteristics of a device based on MoS$_2$ with an intersecting grain boundary. Adapted with permission from ref 95. Copyright 2015 Macmillan Publishers Ltd.: Nature Nanotechnology.

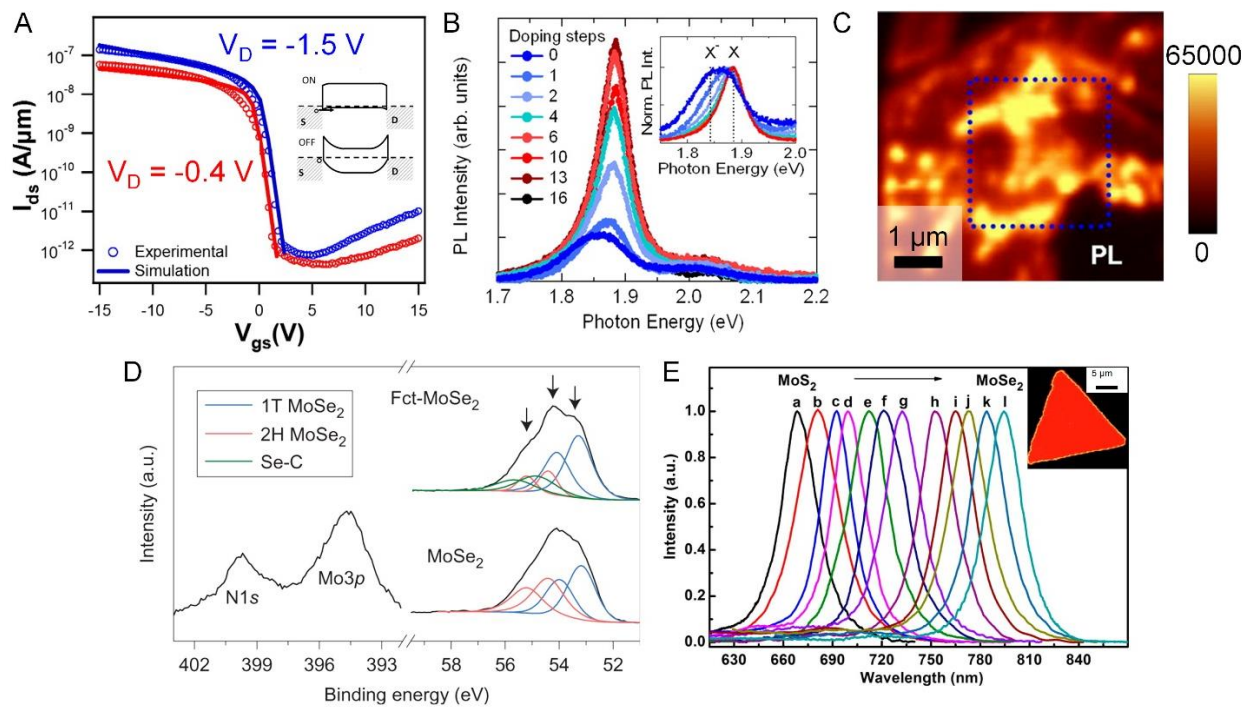

**Figure 3. Chemical manipulation of TMDC optical and electronic properties. (A)** *p*-type MoS$_2$ FET behavior achieved through MoO$_x$ contacts. The inset shows the band levels in the on and off states, where hole injection is enhanced in the on state. Adapted with permission from ref 106. Copyright 2014 American Chemical Society. **(B)** *p*-type doping of MoS$_2$ induced by F$_4$-TCNQ functionalization results in decreasing the electron density and enhanced exciton emission. Adapted with permission from ref 122. Copyright 2013 American Chemical Society. **(C)** Oxygen plasma treatment of MoS$_2$ produces chemisorbed oxygen at Mo sites. The strongly bound oxygen created inhomogeneous PL emission that is enhanced at oxygen-induced defect sites. Adapted with permission from ref 125. Copyright 2014 American Chemical Society. **(D)** Covalent functionalization of the metallic 1T phase of MoSe$_2$ by iodoacetamide modification, which restores the semiconducting 2H phase. Adapted with permission from ref 115. Copyright 2014 Macmillan Publishers Ltd.: Nature Chemistry. **(E)** Continuous tuning of PL emission energy from the MoS$_2$ value of 1.86 eV to the MoSe$_2$ value of 1.56 eV is achieved by alloying in a CVD growth using a temperature gradient. Homogeneous emission is shown in the inset. Adapted with permission from ref 135. Copyright 2014 American Chemical Society.

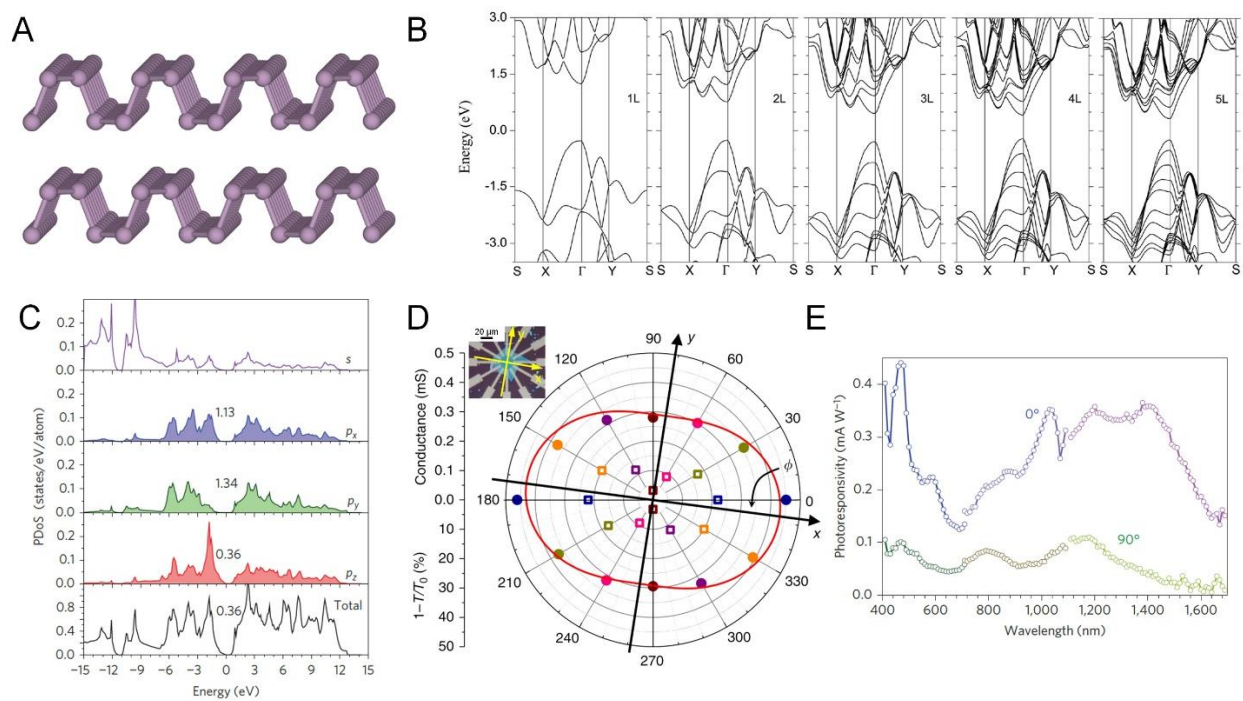

**Figure 4. Black phosphorus (BP) crystal structure, electronic structure, and anisotropy.**
**(A)** Orthorhombic structure of BP. **(B)** DFT calculated band structure of BP for monolayer to five-layer phosphorene, showing a direct band gap magnitude decrease with increasing layer number. Adapted with permission from ref 26. Copyright 2014 Macmillan Publishers Ltd.: Scientific Reports. **(C)** Projected density of states of monolayer black phosphorus showing $s$-$p$ orbital mixing away from the band edges and the strong $p_z$ orbital character of the VBM and CBM. Adapted with permission from ref 159. Copyright 2015 Macmillan Publishers Ltd.: Nature Nanotechnology. **(D)** Anisotropic behavior of BP showing enhanced infrared absorption and conductance in the armchair ($x$) direction. Adapted with permission from ref 144. Copyright 2014 Macmillan Publishers Ltd.: Nature Communications. **(E)** Broadband BP photodetector exhibiting polarization sensitivity in the photoresponse. Adapted with permission from ref 159. Copyright 2015 Macmillan Publishers Ltd.: Nature Nanotechnology.

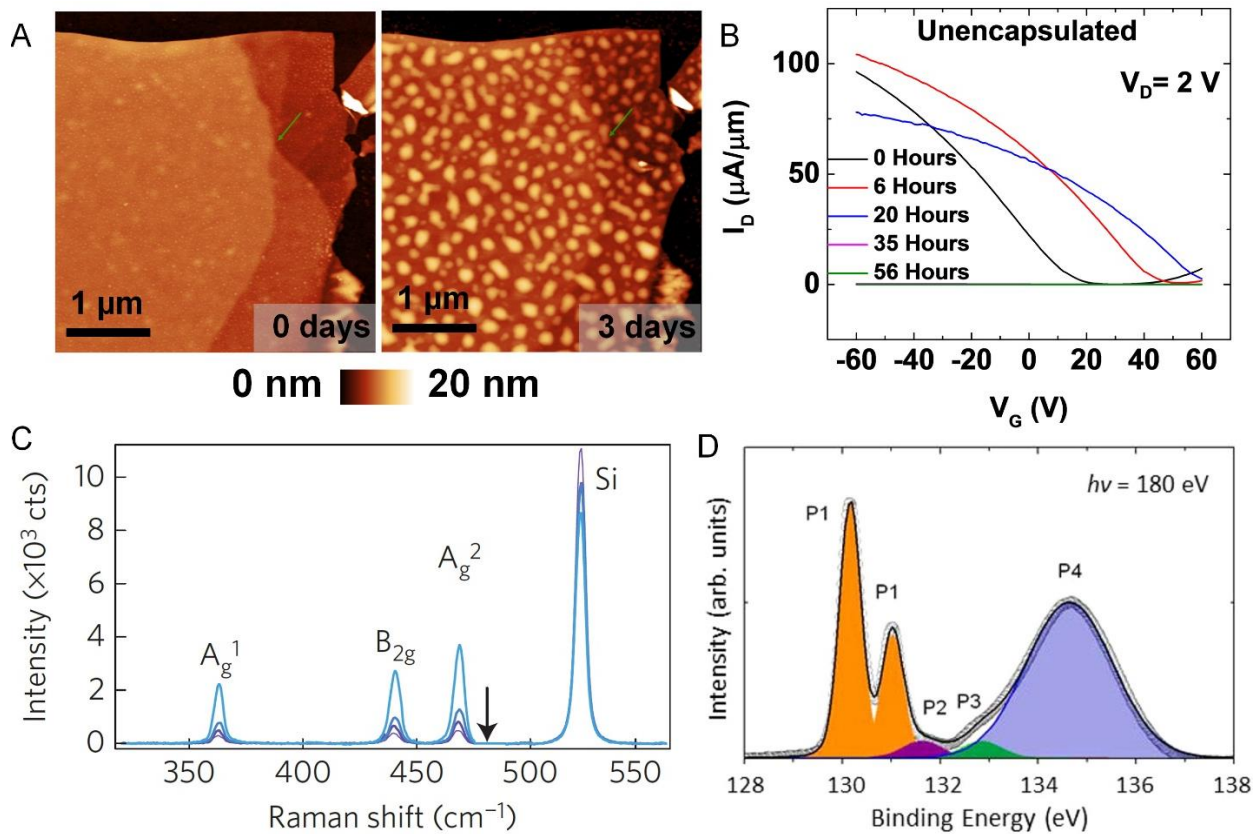

**Figure 5. BP and the effects of ambient environment. (A)** Formation of topographic protrusions in AFM measurements of a BP flake after three days of ambient exposure. Adapted with permission from ref 43. Copyright 2014 American Chemical Society. **(B)** Modest (<20 hours) ambient exposure drives p-type threshold voltage shifts in BP FETs. Over 56 hours of exposure culminates in the loss of channel conductance. Adapted with permission from ref 43. Copyright 2014 American Chemical Society. **(C)** Decrease in BP Raman mode intensity after photo-oxidation in ambient conditions. Adapted with permission from ref 45. Copyright 2015 Macmillan Publishers Ltd.: Nature Materials. **(D)** X-ray photoelectron spectroscopy of oxidized bulk BP, showing the pristine BP doublet ("P1"), $p$-$P_4O_2$ in a bridging state ("P2"), $p$-$P_4O_2$ in a dangling state ("P3"), and layered $P_2O_5$ ("P4"). Adapted with permission from ref 175. Copyright 2015 American Chemical Society.

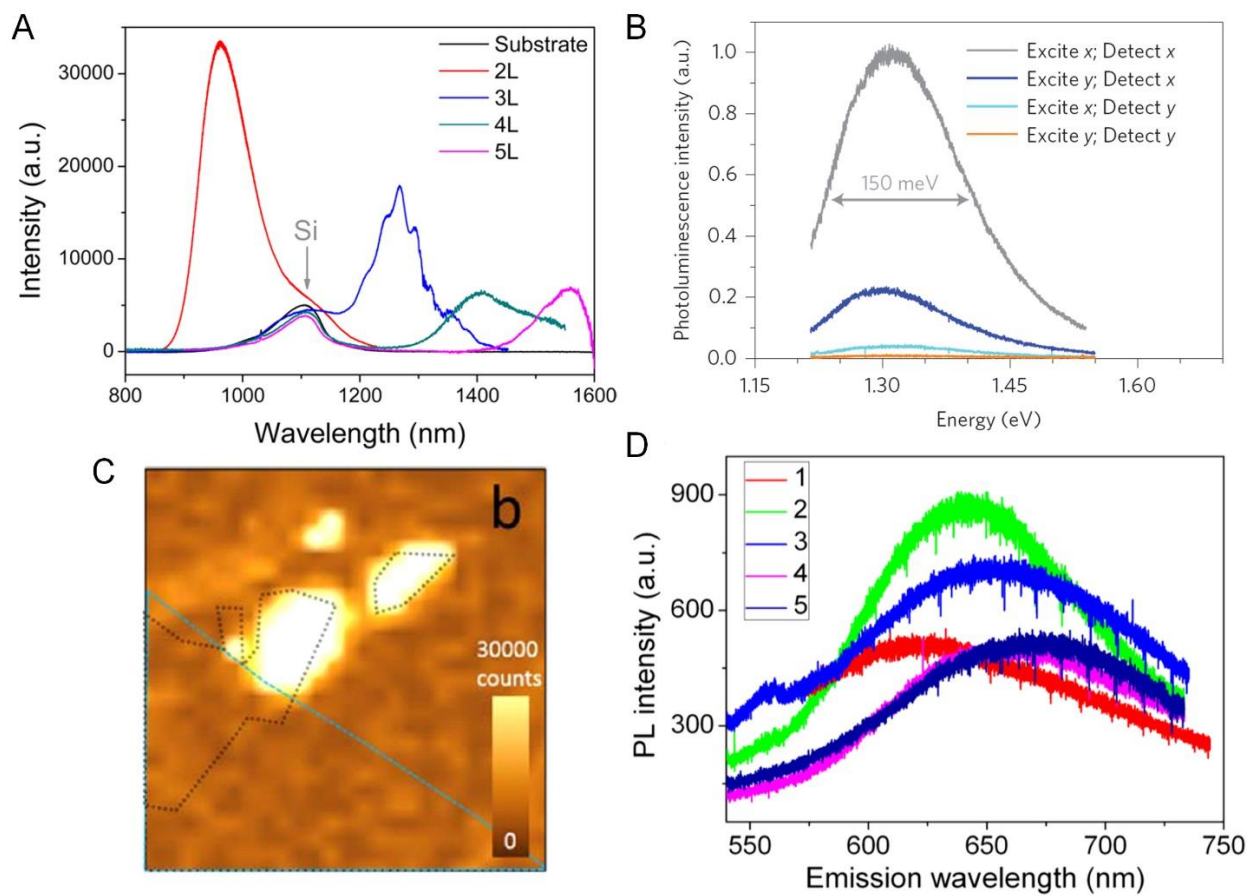

**Figure 6. Chemically dependent BP optical properties. (A)** Photoluminescence (PL) spectra depend on BP layer number. Spectra assigned to 2L (~1.3 eV), 3L (~1.0 eV), 4L (~0.9 eV), and 5L (~0.8 eV). Adapted with permission from ref 183. Copyright 2014 American Chemical Society. **(B)** Anisotropic PL emission at ~1.3 eV, attributed to excitons in monolayer phosphorene. The emission is enhanced by having the excitation and detection polarized parallel to the *x*-direction. Adapted with permission from ref 185. Copyright 2015 Macmillan Publishers Ltd.: Nature Nanotechnology. **(C)** PL map for bilayer phosphorene (black dashes), partly exposed from under a h-BN flake (turquoise triangle). Map intensity is centered about ~1.6 eV, revealing visible BP emission only in ambient exposed regions. Adapted with permission from ref 50. Copyright 2015 American Chemical Society. **(D)** Inhomogeneous PL emission at ~2.0 eV for bulk BP deliberately oxidized by an electrochemical method. Adapted with permission from ref 188. Copyright 2015 AIP Publishing LLC.

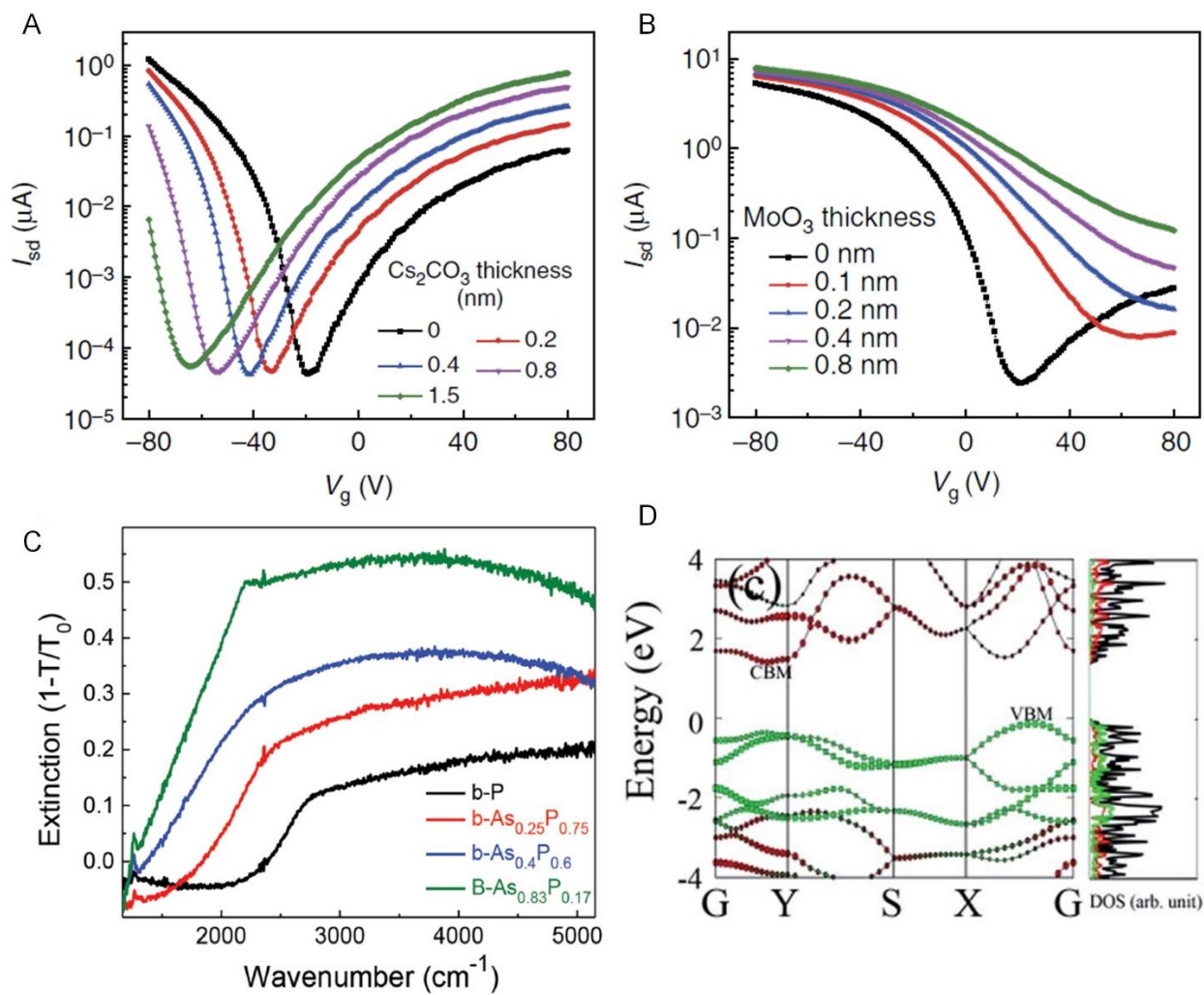

**Figure 7. Chemical manipulation of BP electronic properties. (A)** Continuous *n*-type doping of BP FETs with increasing $Cs_2CO_3$ film thickness. Adapted with permission from ref 189. Copyright 2015 Macmillan Publishers Ltd.: Nature Communications. **(B)** Pronounced *p*-type doping effect produced by $MoO_3$ non-covalent functionalization. Adapted with permission from ref 189. Copyright 2015 Macmillan Publishers Ltd.: Nature Communications. **(C)** Optical band gap modification by alloying BP with arsenic, narrowing of the onset of IR absorption. Adapted with permission from ref 182. Copyright 2015 John Wiley and Sons. **(D)** Band structure and density of states calculations for BP covalently functionalized with $NH_2$. Adapted with permission from ref 177. Copyright 2014 The Royal Society of Chemistry.

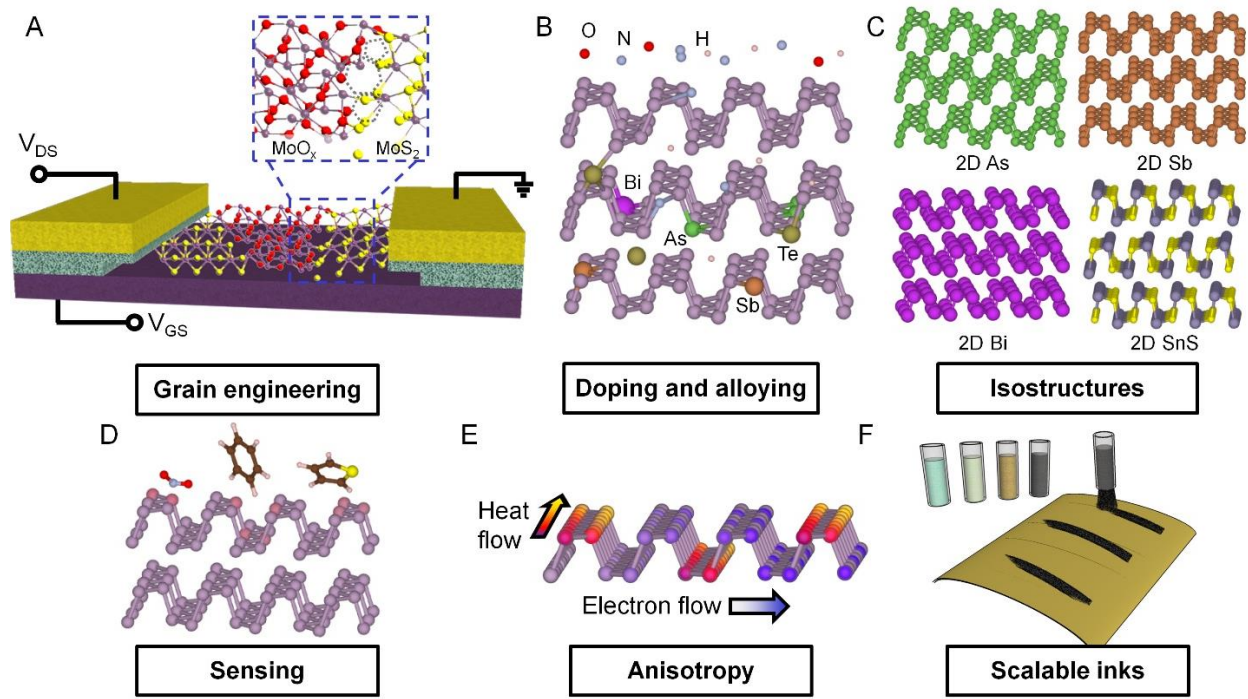

**Figure 8. Outlook for chemically tailored, 2D nanomaterial devices. (A)** TMDC memristors using chemically heterogeneous (*e.g.*, $MoO_x$ and $MoS_2$) grains and grain boundaries. **(B)** Covalent and non-covalent modifications of BP by atomic species (*e.g.*, O, N, and H), doping (*e.g.*, Te), and group 15 alloying (*e.g.*, As, Sb, and Bi). **(C)** Potential elemental and compound isostructures of 2D BP. **(D)** Sensitive and selective sensing (*e.g.*, $NO_2$, benzene, and thiophene) by chemically reactive BP. **(E)** Exploiting BP anisotropy for in-plane thermoelectrics. Thermal conductivity is highest along the in-plane zigzag direction, whereas the electrical conductivity is highest along the armchair direction. **(F)** Wafer-scale BP and group 15 technology through printable, stable inks.